\documentclass[a4paper]{jpconf}
\usepackage{graphicx}
\usepackage{citesort}
\usepackage{amsmath,amssymb}

\begin{document}
\title{Zero-temperature Phase Diagram of Two Dimensional Hubbard Model}

\author{K. Inaba$^a$, A. Koga$^b$, S. Suga$^a$, and N. Kawakami$^b$}
\address{$^a$Department of Applied Physics, Osaka University, Suita, Osaka 565-0871, Japan}
\address{$^b$Department of Physics, Kyoto University, Kyoto 606-8502, Japan}

\ead{inaba@tp.ap.eng.osaka-u.ac.jp}

\begin{abstract}
We investigate the two-dimensional Hubbard model on 
the triangular lattice with anisotropic hopping integrals at half filling.
By means of a self-energy functional approach, 
we discuss how stable the non-magnetic state is 
against magnetically ordered states in the system.
We present the zero-temperature phase diagram, 
where the normal metallic state competes with magnetically 
ordered states with $(\pi, \pi)$ and $(2\pi/3, 2\pi/3)$ structures.
It is shown that a non-magnetic Mott insulating state is not realized 
as the ground state, in the present framework, but as 
a meta-stable state near
the magnetically ordered phase with $(2\pi/3, 2\pi/3)$ structure.
\end{abstract}

Geometrical frustration in
strongly correlated electron systems has attracted current interest.
One of the typical examples is 
an organic material $\kappa$-(BEDT-TTF)$_2$Cu$_2$(CN)$_3$
with the triangular lattice structure, where a non-magnetic insulating state 
is realized down to low temperatures $T\sim$32mK\cite{Shimizu03}.
This suggests that a novel spin-liquid insulating state is stable
against magnetically ordered states,
which stimulates further theoretical investigations on 
frustrated electron systems
\cite{Morita02,Sahebsara06,Kyung06,Mizusaki06,Koretsune07,Watanabe07,Ohashi08,Nevidomskyy08}.
Although a non-magnetic insulating state was shown to be a probable candidate
for the ground state of the Hubbard model on the triangular lattice
\cite{Morita02,Sahebsara06,Kyung06,Koretsune07},
it has not been discussed how this state competes with
an antiferromagnetic insulating state with $120^\circ$ spin structure 
($120^\circ$-AFI) expected naively.
Recently, it was claimed that 
the $120^\circ$-AFI state is, instead of the nonmagnetic insulating state, 
stabilized in the strong coupling regime
by means of variational Monte Carlo (VMC) simulations \cite{Watanabe07}.
However, this method may not deal with 
non-local correlations properly, which should be important
in the two-dimensional frustrated systems.
Therefore, it is highly desirable to deal with intersite correlations 
as well as onsite correlations on an equal footing in order 
 to clarify the ground-state properties of the frustrated systems.

\begin{figure}[t]
\begin{minipage}[b]{.4\linewidth}
\includegraphics[width=.85\linewidth]{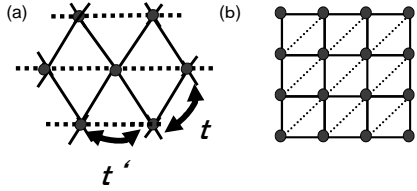}
\end{minipage}
\begin{minipage}[b]{.42\linewidth}
\caption{
(a) [(b)] Sketch of the structure of the model Hamiltonian
(a topologically equivalent structure of the model).
}
\label{fig_original_model}
\end{minipage}
\end{figure}

Motivated by this,
we study the two-dimensional half-filled Hubbard model 
with geometrical frustration.
The model Hamiltonian is given by 
${\cal H}={\cal H}_0({\bf t})+{\cal H}^\prime({\bf U})$ with
\begin{eqnarray}
{\cal H}_0({\bf t})&=&\sum_{{\bf rr}'}\sum_{\sigma} t_{{\bf rr}'} c^\dag_{{\bf r}\sigma} c_{{\bf r}'\sigma}, \\
{\cal H}^\prime({\bf U})&=& \!\!\!\!
                \sum_{\bf r}  U n_{{\bf r} \uparrow} n_{{\bf r} \downarrow}
          \label{eq_twoD_model},
\end{eqnarray}
and ${\bf t}$ (${\bf U}$) is the parameter-matrix of 
the one-particle (two-particles) term, 
where $c^\dag_{{\bf r}\sigma}(c_{{\bf r}\sigma})$ creates (annihilates) 
an electron with spin $\sigma (=\uparrow, \downarrow)$ 
at site ${\bf r}$, 
and $n_{{\bf r}\sigma}$ is the number operator.
Here, $t_{{\bf rr}'}=t (t')$ is 
the nearest-neighbor (second nearest-neighbor) hopping, 
which is schematically shown in Fig. \ref{fig_original_model} (a), 
and $U$ the on-site Coulomb interaction.
We also give a sketch of the square lattice with 
specific diagonal hopping in Fig. \ref{fig_original_model} (b), 
which is topologically 
equivalent to the triangular lattice with anisotropic hopping.
When $t=t'$, the system is equivalent to the Hubbard model on 
the triangular lattice.
In the following, 
we set $t=1$ as the energy unit.

To study the ground-state properties of the system,
we make use of a self-energy functional approach (SFA),
which is based on 
the Luttinger-Ward variational principle \cite{Luttinger}.
It provides us with a non-perturbative treatment of strong correlations, and
an efficient and tractable way to investigate the metal-insulator transition
 (MIT) with and without symmetry breaking
\cite{Potthoff03,Potthoff03L,Dahnken04,Sahebsara06}.
An advantage in this method is to take into account
intersite correlations as well as local electron correlations
on an equal footing, which enables us to discuss the competition between 
the nonmagnetic insulating state and the magnetically ordered states.

In this method, the ground potential $\Omega$ is given as a function of 
a {\it reference} self-energy ${\boldsymbol \Sigma}({\bf t}')$,
\begin{eqnarray}
\Omega [{\boldsymbol \Sigma}({\bf t}^\prime)]&=&\Omega({\bf t^\prime})
    +{\rm Tr}\ln
    \left\{-\left[\omega+\mu-{\bf t}-{\boldsymbol \Sigma}
({\bf t^\prime})\right]^{-1}\right\}
    -{\rm Tr}\ln
    \left\{-\left[\omega+\mu-{\bf t}'-{\boldsymbol \Sigma}
({\bf t^\prime})\right]^{-1}\right\},
\label{eq_twoD_omega_SFA}
\end{eqnarray}
where $\Omega({\bf t}')$ is 
the ground potential for a reference system
with the Hamiltonian ${\cal H}_{ref}={\cal H}_0({\bf t}')+{\cal H}'({\bf U})$.
We note that the Hamiltonian of the reference system 
is defined by the substitution 
of the parameter-matrix ${\bf t}'$ for ${\bf t}$
in the original Hubbard Hamiltonian, where we should keep ${\bf U}$ unchanged.
The variational condition 
$\partial \Omega[{\boldsymbol \Sigma({\bf t}')}]/\partial {\bf t}'=0$
provides us with an approximate self-energy
which properly describes physical properties of the original model.
When one applies the SFA to the present system,
a cluster Anderson impurity model is one of the most appropriate 
reference systems.
It is described by the following Hamiltonian, 
\begin{eqnarray}
  {\cal H}_{0}({\bf t}')&=& \sum_{\bf R}
\bigg\{
  		\sum_{{\bf r}{\bf r}'}\sum_{\sigma} t^{c}_{{\bf rr}'}
          c^\dag_{{\bf r}+{\bf R},\sigma} c_{{\bf r}'+{\bf R},\sigma}
     +\sum_{\ell=1}^{N_b}\sum_{{\bf r}}\sum_{\sigma}  
		\varepsilon^a_{\ell{\bf r}} 
          a^{\dag}_{\ell{\bf r}\sigma}a^{}_{\ell{\bf r}\sigma}\nonumber\\
     &+&\sum_{\ell=1}^{N_b}\sum_{{\bf rr}'}\sum_{\sigma} V^{}_{\ell}
          (c^\dag_{{\bf r}+{\bf R},\sigma}a^{}_{\ell{\bf r}'\sigma}+H.c.)\nonumber
     + \sum_{{\bf r}} \exp(i {\bf r}\cdot{\bf Q}){\bf H}_{\bf Q}\cdot{\bf S}_{{\bf r}+{\bf R}}\bigg\},
\label{eq_twoD_ref_model}
\end{eqnarray}
where $\sum_{\bf r}$ ($\sum_{\bf R}$) sums up sites (clusters) 
in the cluster (the whole lattice) and $\{ t^c_{\bf rr'}, \varepsilon_{\ell{\bf r}}^a, V_\ell,{\bf H_Q}\}$ are variational parameters.
$a^{\dag}_{\ell{\bf r}\sigma}(a^{}_{\ell{\bf r}\sigma})$
creates (annihilates) an electron with spin $\sigma$ 
at the $\ell(=1, \cdots, N_b)$th site in the effective bath,
which is connected to the original lattice at site {\bf r}.
In order to deal with magnetically ordered states,
we introduce the effective magnetic field ${\bf H_Q}$ 
in the reference system.
Here, ${\bf S}_{\bf r}=\frac{1}{2}c_{\bf r\gamma}^\dag 
{\bf \sigma}_{\gamma\gamma'} 
c_{\bf r\gamma'}$, where ${\bf \sigma}$ is the Pauli matrix.
In this study, we mainly discuss the triangular lattice, $t=t'$. 
In the following, 
to investigate the MIT of the half-filled Hubbard model,
we examine a free energy $F=\Omega - \mu N$ with
an additional variational condition $\partial F/\partial \mu = 0$, 
where $\mu$ is the chemical potential and $N$ is the total number of electrons. 
\begin{figure}[t]
\begin{center}
\includegraphics[width=.34\linewidth]{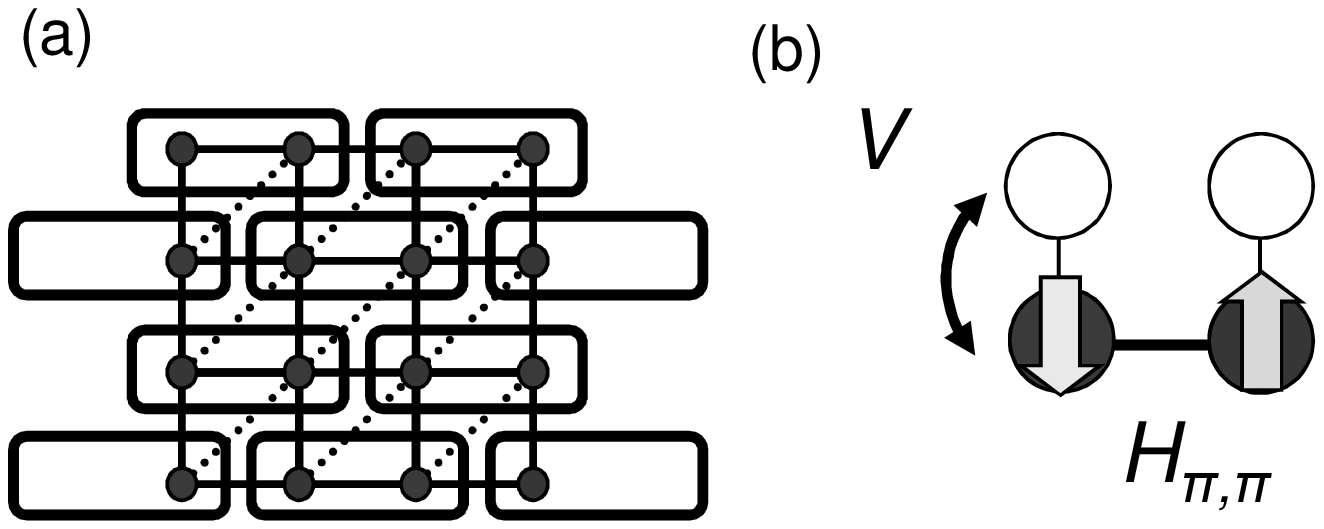}\hspace{2cm}
\includegraphics[width=.33\linewidth]{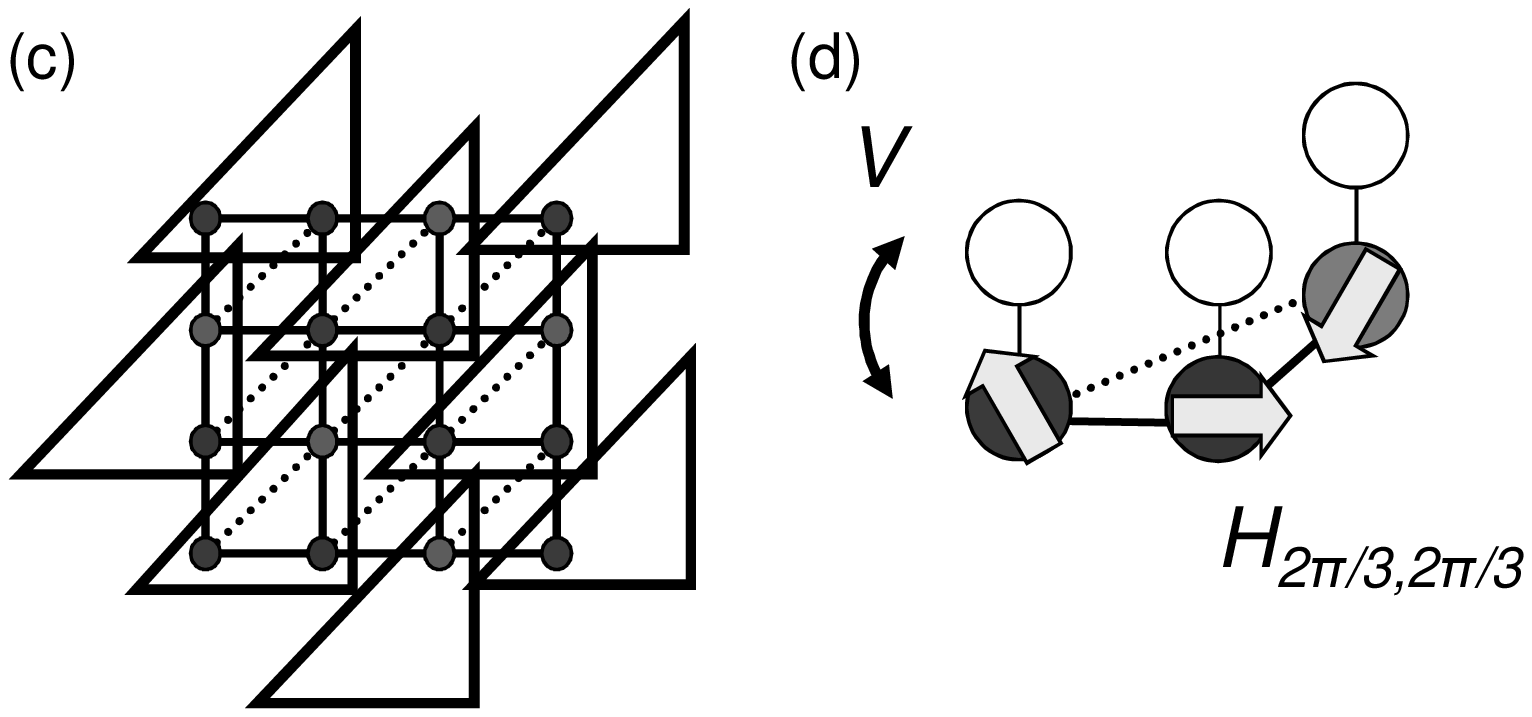}
\end{center}
\caption{
(a) [(c)] Lattice structure for two-(three-)site clusters, 
which are represented by rectangles (triangles).
(b) [(d)] Cluster impurity model to describe 
a magnetically ordered state in the original model, 
where arrows indicate the direction of the effective magnetic field.
Dark (white) circles correspond to the impurity-(bath-)site.
}\label{fig_ref_model}
\end{figure}

\begin{figure}[t]
\begin{center}
\includegraphics[width=.36\linewidth]{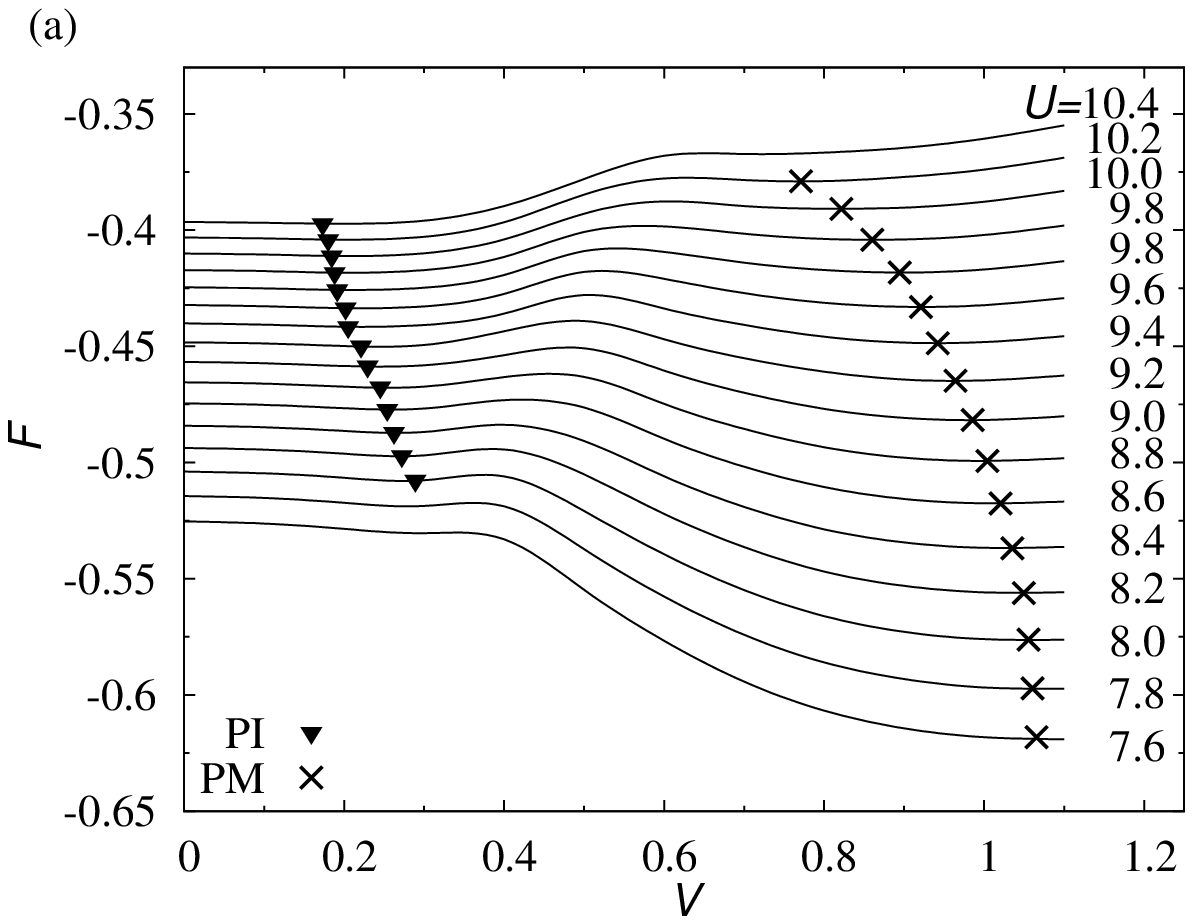}
\hspace{1cm}
\includegraphics[width=.36\linewidth]{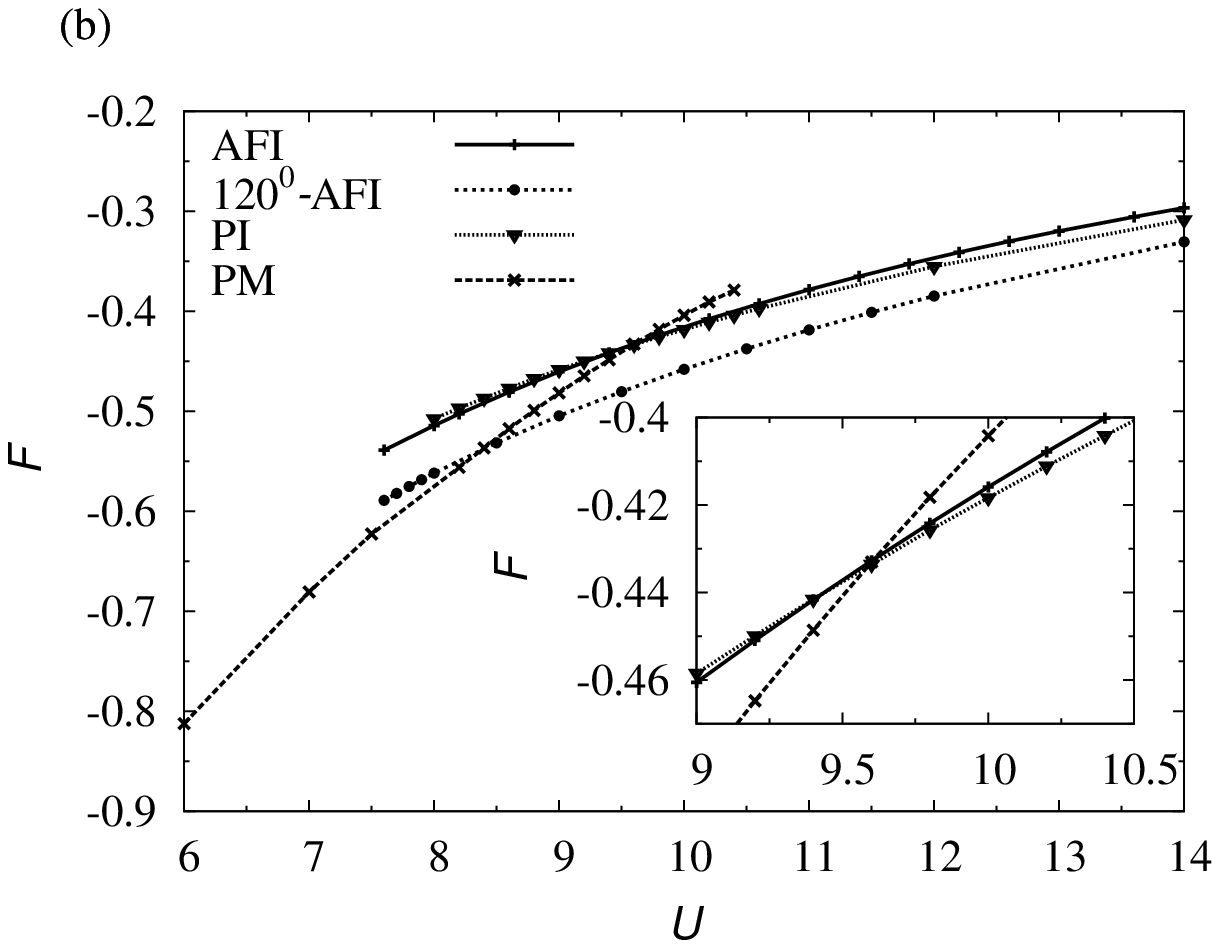}
\end{center}
\caption{(a) Free energy $F$ as a function of 
the variational parameter $V$ for $t'=1$.
Triangles (crosses) correspond to the stationary points for 
the paramagnetic insulating (paramagnetic metallic) states.
(b) Free energies $F$ for several states as a function of $U$ when $t'=1$. 
The inset shows an enlarged picture around $U\sim9.5$.
}\label{fig_free_energy}
\end{figure}
Before addressing a magnetic instability,
we first consider the zero-temperature properties in the paramagnetic state.
This might be important at very low temperatures
since magnetically ordered states become unstable in two dimensions.
Here, we use a two-site cluster with ${\bf H_Q}=0$ 
as a reference system, which is
illustrated in Fig. \ref{fig_ref_model} (b).
The obtained free energy is shown in Fig. \ref{fig_free_energy} (a).
In the small $U$ region  ($U\lesssim 7.8$),
we find one stationary point with a large $V$, 
which should represent the paramagnetic metallic (PM) state.
The increase in $U$ gradually changes the stationary point and 
decreases its variational parameter $V$, 
which implies that the heavy quasi-particle state is realized in the case $U\lesssim 7.8$.
On the other hand, the paramagnetic insulating (PI) state appears 
in the strong coupling region ($U\gtrsim 10.3$),
which is shown as another stationary point in the figure.
Thus, these two phases compete with each other and 
a first-order MIT between the PI and the PM states occurs around $U\sim 9.6$
as far as the paramagnetic states are concerned.

However, this result does not necessarily imply that 
the PI state is stable against magnetically ordered states 
at zero temperature.
To make this clear, we discuss the magnetic instability in the system.
Potential candidates for the ordered states are the 
($\pi,\pi$)-antiferromagnetic insulating (AFI) state and
the $120^\circ$-AFI state.
By using two kinds of the reference systems 
illustrated in Figs. \ref{fig_ref_model} (b) and (d),
we obtain the free energies for paramagnetic and magnetically ordered states, 
as shown in Fig. \ref{fig_free_energy} (b).
We can see that the PM state is stable in the small $U$ region.
On the other hand, it is found that in the strong coupling limit, 
the $120^\circ$-AFI state is stabilized, 
which is consistent with the results for the Heisenberg model.
At $U_c\sim 8.4$, 
the free energies for these states cross each other, 
 suggesting the first-order MIT accompanied by symmetry breaking.
We also find that the non-magnetic insulating state 
is no longer realized as the ground state within the present approach.
Nevertheless in the PM region ($7.7 \lesssim U \lesssim 8.4$),
the non-magnetic insulating state exists as a meta-stable state.
Therefore, 
if the interacting system is adiabatically cooled down,
the  meta-stable non-magnetic insulating state could emerge
down to zero temperature.

\begin{figure}[t]
\begin{minipage}[b]{.42\linewidth}
\includegraphics[width=.9\linewidth]{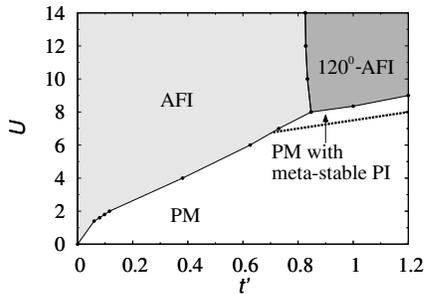}
\end{minipage}
\begin{minipage}[b]{.52\linewidth}
\caption{\label{fig_phase}
Phase diagram of the half-filled Hubbard model.
White region represents the PM phase, 
and light (dark) gray region the ($\pi,\pi$)-AFI 
[$120^\circ$-AFI] phase.
The thick dashed line indicates the boundary 
where the meta-stable PI state disappears with decreasing $U$.
}
\end{minipage}
\end{figure}

By performing SFA calculations with several values of $t'$,
we obtain the ground-state phase diagram 
of the half-filled Hubbard model, 
which is shown in Fig. \ref{fig_phase}.
The PM state is stable in the weakly correlated region with $t'\neq 0$.
In the strongly correlated region,
the ($\pi, \pi$)-AFI state competes with the $120^\circ$-AFI state 
and the phase transition occurs at $t_c'\sim 0.8$.
These results are 
consistent with those obtained by the VMC calculations \cite{Watanabe07}.
 In Fig. \ref{fig_phase}, we show the region where
the non-magnetic insulating state is not realized as the ground state, 
but appears as a  meta-stable state.

In summary, 
we have investigated 
the half-filled Hubbard model on 
the triangular lattice with anisotropic hopping integrals
by means of a self-energy functional approach.
We have clarified how the normal metallic state competes with the magnetically 
ordered states with $(\pi, \pi)$ and $(2\pi/3, 2\pi/3)$ structures.
When $t\neq t'$, another type of magnetically ordered state with 
an incommensurate wave vector ${\bf Q}$ might be also possible
\cite{Weihong99},
which has not been addressed in this paper.
Therefore, it remains an important problem 
to explore the stability of such 
ordered states,
which is now under consideration.
Furthermore, it has been suggested that 
in the square-lattice model with next-nearest hopping, 
which is another important frustrated system slightly different from the present one,
the collinear ordered phase with $(\pi, 0)$ or $(0, \pi)$ 
is stabilized for large frustration\cite{Mizusaki06,Nevidomskyy08}. 
Comprehensive investigations 
including such frustrated systems are important, 
which are left for the future study.

Numerical computations were carried out at the Supercomputer Center, the Institute for Solid State Physics, University of Tokyo. KI was supported by the Japan Society for the Promotion of Science. This work was supported by a Grant-in-Aid for Scientific Research from the Ministry of Education, Culture, Sports, Science and Technology, Japan (No. 20029013, No. 19014013, No. 20740194).
\section*{References}
\providecommand{\newblock}{}

\end{document}